\documentclass[prd,twocolumn,showpacs,floatfix,amsmath,nofootinbib,amssymb,floatfix]{revtex4}
\usepackage{graphicx,color,dcolumn,booktabs,bm}
\usepackage{longtable,lscape}
\usepackage{txfonts}
\usepackage{overpic}
\usepackage{amssymb}
\usepackage{indentfirst}
\usepackage{feynmf}   %{feynmp}
\usepackage{slashed}  %for Feynman symbols
\usepackage{cases}
\usepackage{color}
\usepackage{multirow}
\usepackage{epstopdf}
\usepackage{graphicx,color,dcolumn,booktabs,bm}
\usepackage[colorlinks, citecolor=blue,anchorcolor=red,menucolor=red, linkcolor=red,filecolor=red,runcolor=red,urlcolor=blue,frenchlinks=red]{hyperref}

\def\lrpartial{\buildrel\leftrightarrow\over\partial}
\graphicspath{{Figures/}} %

\begin{document}
%\begin{CJK}{GBK}{}

\title{Observation of $e^+e^-\to \chi_{c0}\omega$ and missing higher charmonium $\psi(4S)$}
\author{Dian-Yong Chen$^{1,2}$}\email{chendy@impcas.ac.cn}
\author{Xiang Liu$^{2,3}$\footnote{Corresponding author}}\email{xiangliu@lzu.edu.cn}
\author{Takayuki Matsuki$^{4,5}$}\email{matsuki@tokyo-kasei.ac.jp}
\affiliation{$^1$Nuclear Theory Group, Institute of Modern Physics, Chinese Academy of Sciences, Lanzhou 730000, China\\
$^2$Research Center for Hadron and CSR Physics,
Lanzhou University $\&$ Institute of Modern Physics of CAS,
Lanzhou 730000, China\\
$^3$School of Physical Science and Technology, Lanzhou University,
Lanzhou 730000, China\\
$^4$Tokyo Kasei University, 1-18-1 Kaga, Itabashi, Tokyo 173-8602, Japan\\
$^5$Theoretical Research Division, Nishina Center, RIKEN, Saitama 351-0198, Japan}

\begin{abstract}
Stimulated by the recent BESIII observation of a new resonance in $e^+ e^- \to \omega \chi_{c0}$ which is consistent with our predicted $\psi(4S)$, we estimate the meson loop contribution to $\psi(4S) \to \omega \chi_{c0}$ in this work. The evaluation indicates that our theoretical estimate can overlap with the experimental data in a reasonable parameter range. This fact shows that introduction of the missing higher charmonium $\psi(4S)$ provides a possible explanation to the recent BESIII observation. The upper limit of a branching ratio of $\psi(4S) \to \eta J/\psi$ is also predicted to be $1.9 \times 10^{-3}$, which can be further tested by BESIII, Belle and forthcoming BelleII.
\end{abstract}

\pacs{14.40.Pq, 13.66.Bc}

\maketitle

\section{introduction}\label{sec1}

Although the present $J/\psi$ family has become more and more abundant, there still exist much more puzzling features of charmonia, especially higher than 4 GeV, which are waiting for resolution. In the past decade, experiments have made big progress on searching for the charmonium-like states normally referred to $XYZ$, where a lot of them come from the $e^+e^-$ annihilation processes, e.g., $Y(4260)$ observed in $e^+e^-\to J/\psi\pi^+\pi^-$ \cite{Aubert:2005rm}, $Y(4360)$ \cite{Aubert:2007zz} and $Y(4660)$ \cite{Wang:2007ea} reported in $e^+e^-\to \psi(2S)\pi^+\pi^-$, and $Y(4630)$ existing in $e^+e^-\to \Lambda_c^+\bar{\Lambda}_c^-$ \cite{Pakhlova:2008vn}. Thus, the $e^+e^-$ annihilation process is a good platform to explore charmonium-like states.  By carrying out the study of these experimental observations, which have a close relationship with higher charmonia above 4 GeV, it is helpful to establish the $J/\psi$ family and to definitely identify exotic states. The underlying motivation of this study is to enlarge our knowledge of non-perturbative behavior of quantum chromodynamics (QCD), which is a crucial step to gain a deeper understanding of strong interaction.

Very recently, the BESIII Collaboration announced the observation of enhancement in the process $e^+e^-\to \omega\chi_{c0}$ after performing the search for $e^+e^-\to \omega\chi_{cJ}$ $(J=0,1,2)$, which is based on the collected data by BESIII at nine center-of-mass energies from 4.21 to 4.42 GeV \cite{Ablikim:2014qwy}. Among the measured Born cross sections at nine energy points, those at $\sqrt{s}=4.23$ GeV and $4.26$ GeV are $(55.4\pm6.0\pm5.9)$ pb and $(23.7\pm5.3\pm3.5)$ pb, respectively.
There are, however, no significant signals for the remaining processes $e^+e^-\to \omega\chi_{c1}$ and $e^+e^-\to \omega\chi_{c2}$. When one uses the Breit-Wigner function to fit the experimental data of $e^+e^-\to \omega\chi_{c0}$, a resonance structure with mass $M=(4230\pm 8)$ MeV and width $\Gamma=(38\pm12)$ MeV was observed \cite{Ablikim:2014qwy}. BESIII indicated that this resonance structure is different from $Y(4260)$ reported in the analysis of $e^+e^-\to J/\psi\pi^+\pi^-$ \cite{Aubert:2005rm}. It is a challengeable and intriguing task how to understand this novel phenomenon.

To reveal the underlying mechanism behind the above observation, the enhancement structure around 4230 MeV with a narrow width can provide a valuable hint and give an answer why only $e^+e^-\to \omega\chi_{c0}$ was observed in BESIII.
Before this BESIII observation, we have once predicted a missing higher charmonium $\psi(4S)$ with the mass 4263 MeV and narrow width utilizing the similarity between the $J/\psi$ and $\Upsilon$ families \cite{He:2014xna}.
{There are some theoretical calculations \cite{Barnes:2007xu,Segovia:2008zz,Li:2009ad} of the mass spectrum of the charmonium by considering the coupled-channel effect. In fact, the present calculation if including the coupled-channel effect is model-dependent since different coupled-channel models have given different results of the mass spectrum of the charmonium family \cite{Barnes:2007xu,Segovia:2008zz,Li:2009ad}. To support our prediction of a missing $\psi(4S)$ around 4.2 GeV, we notice three theoretical papers \cite{Ding:1993uy,Dong:1994zj,Li:2009zu}, where the screened potential is considered. Their calculations of $\psi(4S)$ support a $\psi(4S)$ with mass around 4.2 GeV, i.e., 4273 MeV \cite{Li:2009zu} and 4247 MeV \cite{Dong:1994zj}. To some extent, introduction of the screened potential is an equivalent description of the coupled-channel effect, which was indicated in Ref. \cite{Li:2009ad}. }

Comparing the resonance parameters of the predicted $\psi(4S)$ with those of the enhancement structure given by BESIII \cite{Aubert:2005rm}, we notice the structure existing in $e^+e^-\to \omega\chi_{c0}$ is consistent with
our predicted missing $\psi(4S)$. Therefore, the enhancement around 4230 MeV in $e^+e^-\to  \omega\chi_{c0}$ can be identified as $\psi(4S)$. {With this assignment, we need to further explain the cross section of $e^+ e^- \to \omega \chi_{c0}$ experimentally measured, which is the main task of this work.} Since the threshold values of $\omega\chi_{c0}$, $\omega\chi_{c1}$, and $\omega\chi_{c2}$ are 4197 MeV, 4293 MeV, and 4338 MeV, respectively, it is obvious that the central mass of $\psi(4S)$ (if taking the experimental value $M=4230\pm8$ MeV \cite{Ablikim:2014qwy}) is just above the $\omega\chi_{c0}$ threshold and below the $\omega\chi_{c1,2}$ thresholds. This fact naturally explains why only $e^+e^-\to \omega\chi_{c0}$ was observed for the first time in BESIII. This is because introduction of a long-term missing $\psi(4S)$ kinematically forbids $\psi(4S)\to \omega\chi_{c1,2}$.

In the following, we will study the $e^+e^-\to \omega\chi_{c0}$ process via the intermediate $\psi(4S)$.
Since the Born cross sections of $e^+e^-\to \omega\chi_{c0}$ were measured by BESIII \cite{Ablikim:2014qwy}, we are able to compare our numerical results with the experimental data, which can be tested whether introduction of $\psi(4S)$ contribution is reasonable to explain this recent BESIII observation. In the next section, we present more details of the calculation of $e^+e^-\to \psi(4S) \to \omega\chi_{c0}$.

The decay $\psi(4S)\to \eta J/\psi$ similar to $\psi(4S) \to \omega\chi_{c0}$ can occur, which is a typical transition accessible by experiment. Hence, in this work we also study the $\psi(4S)\to \eta J/\psi$ decay, where the partial width of this decacy and the cross section of $e^+e^-\to \psi(4S) \to \eta J/\psi$ are predicted. These are important informations for experimentalists to further search for the $e^+e^-\to \eta J/\psi$ process, which can be seen as a further test of our understanding of the BESIII observation of $e^+e^-\to \omega\chi_{c0}$.

This work is organized as follows. After Introduction, we present the detailed calculation of $\psi(4S)\to \omega\chi_{c0}$ and $\psi(4S)\to \eta J/\psi$ in Sec. \ref{sec2}. In Sec. \ref{sec3}, the numerical results are given. The last section ends with the conclusions and discussion.

\section{$\psi(4S)\to \omega\chi_{c0}$ and $\psi(4S)\to \eta J/\psi$ transitions}\label{sec2}

For the hidden-charm decays of higher charmonium, the hadronic loop mechanism plays an important role to mediate these decays. In the past decade, there were some discussions of this point, which shows that the novel phenomena existing in the decays of higher charmonia, bottomonia and $B$ mesons can be indeed understood well by the hadronic loop mechanism \cite{Liu:2006dq,Li:2007au,Liu:2009dr,Guo:2009wr,Zhang:2009kr,Chen:2009ah,Chen:2012nva,Chen:2014ccr,Li:2007xr,Chen:2010re,Chen:2013cpa,Colangelo:2003sa,Colangelo:2002mj}.

\begin{center}
\begin{figure}[htb]
\scalebox{0.5}{\includegraphics{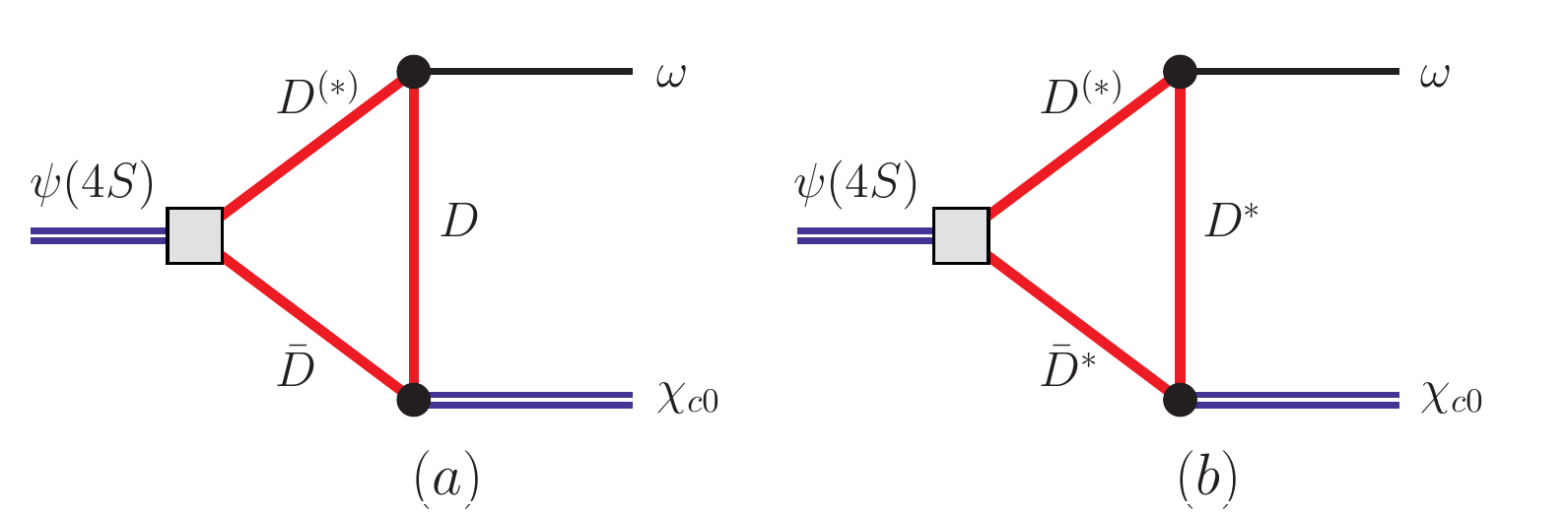}}
\caption{Sketchy diagrams of the meson loop contributions to $\psi(4S) \to \omega \chi_{c0}$.
(a) and (b) correspond to $\psi(4S) \to [D^{(*)}\bar{D}]_{D} \to \omega \chi_{c0}$ and $\psi(4S) \to [D^{(*)}\bar{D}^*]_{D^*}\to \omega \chi_{c0}$ decays, respectively,  \label{Fig:Tri-omegachic0}}
\end{figure}
\end{center}

For the discussed $\psi(4S)\to \omega\chi_{c0}$ and $\psi(4S)\to \eta J/\psi$ transitions, $\psi(4S)$ is above the threshold of $D_{(s)}^{(\ast)} \bar{D}_{(s)}^{(\ast)}$ and dominantly decays into a pair of charmed mesons or charmed strange mesons \cite{He:2014xna}, where the corresponding partial decay widths were calculated in Ref. \cite{He:2014xna}. On the other hand, the $\psi(4S)\to \omega\chi_{c0}$ and $\psi(4S)\to \eta J/\psi$ transitions can occur via the hadronic loop mechanism. For example, the initial state $\psi(4S)$ can decay into the final state $\omega \chi_{c0}$ via the charmed meson loops, which are shown in Fig. \ref{Fig:Tri-omegachic0}.

To calculate the hadronic loop contributions to the $\psi(4S)\to \omega\chi_{c0}$ and $\psi(4S)\to \eta J/\psi$ decays, we utilize the effective Lagrangian approach, where the the effective interaction Lagrangians are constructed by respecting heavy quark limit and chiral symmetry \cite{Kaymakcalan:1983qq, Oh2000qr, Casalbuoni1996pg,
Colangelo2002mj}. The involved effective Lagrangians that are related to interactions among charmonium and charmed mesons and among vector/pseudoscalar meson and charmed mesons read as
\begin{eqnarray}
\mathcal{L}_{\psi \mathcal{D}^{(\ast)} \mathcal{D}^{(\ast)}}
&=& -ig_{\psi \mathcal{DD} } \psi_\mu (\partial^\mu
\mathcal{D} \mathcal{D}^\dagger- \mathcal{D}
\partial^\mu \mathcal{D}^\dagger) \nonumber\\
&& + g_{\psi
\mathcal{D}^\ast \mathcal{D}} \varepsilon^{\mu \nu \alpha \beta}
\partial_\mu \psi_\nu (\mathcal{D}^\ast_\alpha \lrpartial_\beta
\mathcal{D}^\dagger -\mathcal{D} \lrpartial_\beta
\mathcal{D}_\alpha^{\ast \dagger} ) \nonumber\\
&& + ig_{\psi
\mathcal{D}^\ast \mathcal{D}^\ast} \psi^\mu
(\mathcal{D}^\ast_\nu \partial^\nu \mathcal{D}^{\ast \dagger}_\mu
-\partial^\nu \mathcal{D}^{\ast}_\mu \mathcal{D}^{\ast \dagger}_\nu
\nonumber\\&&-\mathcal{D}^\ast_\nu \lrpartial_\mu \mathcal{D}^{\ast \nu
\dagger}), \label{Eq:LpsiDD}%\\
\end{eqnarray}
\begin{eqnarray}
\mathcal{L}_{\chi_{c0} \mathcal{D}^{(\ast)} \mathcal{D}^{(\ast)}}
&=& - g_{\chi_{c0} \mathcal{D} \mathcal{D} } \chi_{c0} \mathcal{D}
\mathcal{D}^\dagger - g_{\chi_{c0} \mathcal{D}^\ast
\mathcal{D}^\ast} \chi_{c0} \mathcal{D}_{\mu}^\ast \mathcal{D}^{\ast
\mu\dagger }, \label{Eq:Lchic0DD}%\\
\end{eqnarray}
\begin{eqnarray}
\mathcal{L}_{\mathcal{D}^{(\ast)}\mathcal{D}^{(\ast)} \mathcal{V}}
&=& -ig_{\mathcal{D} \mathcal{D}\mathcal{V}} \mathcal{D}_i^\dagger
\lrpartial^\mu \mathcal{D}^j (\mathcal{V}_\mu)^i_j -2
f_{\mathcal{D}^\ast \mathcal{D} \mathcal{V}} \varepsilon_{\mu \nu
\alpha \beta}  \nonumber\\&& \times (\partial^\mu
\mathcal{V}^\nu)^i_j (\mathcal{D}^\dagger_i \lrpartial^\alpha
\mathcal{D}^{\ast \beta j} -\mathcal{D}_i^{\ast \beta \dagger}
\lrpartial^\alpha \mathcal{D}^j) \nonumber\\&& +ig_{\mathcal{D}^\ast
\mathcal{D}^\ast \mathcal{V}} \mathcal{D}^{\ast \nu \dagger}_i
\lrpartial^\mu \mathcal{D}^{\ast j}_\nu (\mathcal{V}_\mu)^i_j
\nonumber\\ &&+4if_{\mathcal{D}^\ast \mathcal{D}^\ast \mathcal{V}}
\mathcal{D}^{\ast \dagger}_{i\mu} (\partial^\mu \mathcal{V}^\nu
-\partial^\nu \mathcal{V}^\mu)^i_j \mathcal{D}^{\ast j}_\nu,
\label{Eq:LDDV}
\end{eqnarray}
\begin{eqnarray}
\mathcal{L}_{\mathcal{D}^{(\ast)}\mathcal{D}^{(\ast)}\mathcal{P}}
&=& -i g_{\mathcal{D}^*\mathcal{D}P} (\bar{\mathcal{D}}  \partial_\mu \mathcal{P}
\mathcal{D}^{*\mu}  - \bar{\mathcal{D}}^{*\mu}  \partial_\mu
\mathcal{P}  \mathcal{D} ) \nonumber\\
&& + \frac{1}{2}
g_{\mathcal{D}^*\mathcal{D}^*P}\epsilon_{\mu\nu\alpha\beta} \bar{\mathcal{D}}^{*\mu}
\partial^\nu \mathcal{P}\;  {\stackrel{\leftrightarrow}{\partial^\alpha}}\;
\mathcal{D}^{*\beta} ,\label{Eq:LDDP}
\end{eqnarray}
where $\mathcal{V}$ and $\mathcal{P}$ denote the matrix of the vector octet and pseudoscalar octet, respectively. The explicit expressions for $\mathcal{V}$ and $\mathcal{P}$ are
\begin{eqnarray}
\mathcal{V} &=&
 \left(
 \begin{array}{ccc}
\frac{1}{\sqrt{2}} (\rho^{0}+ \omega) & \rho^{+} & K^{*+}\\
\rho^{-} & \frac{1}{\sqrt{2}}(-\rho^0 +\omega) &  K^{*0}\\
 K^{*-} & \bar{K}^{*0} & \phi
 \end{array}
 \right).
\end{eqnarray}
\begin{eqnarray}
  \mathcal{P} &=&
 \left(
 \begin{array}{ccc}
\frac{\pi^0}{\sqrt{2}} + \alpha \eta + \beta \eta^\prime & \pi^{+} & K^{+}\\
\pi^{-} & -\frac{\pi^0}{\sqrt{2}}+ \alpha \eta + \beta \eta^\prime   &  K^{0}\\
 K^{-} & \bar{K}^{0} & \gamma \eta + \delta \eta^\prime
 \end{array}
 \right),
\end{eqnarray}
where the corresponding mixing angles are defined as
$\alpha=({\cos \theta -\sqrt{2} \sin \theta})/{\sqrt{6}}$, $\beta =
({\sin \theta + \sqrt{2} \cos \theta})/{\sqrt{6}}$, $\gamma = ({-2 \cos \theta -\sqrt{2} \sin \theta})/{\sqrt{6}}$,
and $\delta= ({-2 \sin \theta + \sqrt{2} \cos \theta})/{\sqrt{6}}$.
%\end{eqnarray}
In the present calculations, we adopt $\theta=-19.1^\circ$ determined in Refs. \cite{Coffman:1988ve,Jousset:1988ni}.
Since $\psi(4S)$ and $J/\psi$ have the same $J^{PC}$ quantum numbers, the interaction Lagrangians for $\psi(4S)D^{(\ast)}D^{(\ast)}$ have the same forms as those describing the interactions for $J/\psi D^{(\ast)}D^{(\ast)}$, where the coupling constants will be given later.

With these effective Lagrangians listed in Eqs. (\ref{Eq:LpsiDD})-(\ref{Eq:LDDP}), we can obtain the Feynman rules, which are collected in Appendix.

With the above preparation, we can easily write out the decay amplitudes for $\psi(4S) \to \omega \chi_{c0}$ and
$\psi(4S) \to \eta J/\psi$, where a general form of the decay amplitude is
\begin{eqnarray}
\mathcal{M}=\int \frac{d^4q}{(2\pi)^4}\frac{\mathcal{V}_1\mathcal{V}_2\mathcal{V}_3}{\mathcal{P}_1 \mathcal{P}_2 \mathcal{P}_{E}}\mathcal{F}^2(q, m_E),\label{h2}
\end{eqnarray}
where $\mathcal{V}_{i}$  are triple couplings % corresponding to Eqs. (\ref{Eq:VpsiDD})-(\ref{h1})
and $1/\mathcal{P}_i$ correspond to the propagators
% defined in Eq. (\ref{Eq:propD})-(\ref{Eq:propDStar})
with $1/\mathcal{P}_E$ expressing the exchanged-meson propagator.
The concrete expressions for the amplitude $\psi(4S)\to \omega\chi_{c0}$ with the meson loop contributions are similar to those in our former work relevant to $\Upsilon(5S)\to \omega\chi_{b0}$ \cite{Chen:2014ccr}. In Eq. (\ref{h2}), the form factor $\mathcal{F}(q, m_E)$ is taken as a monopole expression $\mathcal{F}(q, m_E) = (m_E^2-\Lambda^2)/(q^2-\Lambda^2)$,
which is introduced to describe the structure effect of interacting vertices and off-shell effect that results from the exchanged meson. Here, the cutoff $\Lambda$ can be further parameterized as $\Lambda=\alpha_{\Lambda} \Lambda_{QCD} +m_E$ with $\Lambda_{QCD}=220$ MeV and $m_E$ denotes the exchanged-meson mass. {The free parameter $\alpha_{\Lambda}$ should be of order one, which is not a universal parameter and is dependent on the concrete processes \cite{Cheng:2004ru}}. In addition, the introduced form factor plays a role of regularization to get rid of the UV divergence of the loop integrals, which is similar to the Pauli-Villas regularization scheme.

In the following, we elucidate how to determine the values of the involved coupling constants in our calculation.
With the Lagrangian listed in Eq. (\ref{Eq:LpsiDD}), we get the partial decay widths of the open charm decays of $\psi(4S)$, i.e.,
\begin{eqnarray}
\Gamma_{\psi(4S) \to DD} &=& \frac{g_{\psi(4S)DD}^2 \lambda(m_{\psi(4S)}^2,m_{D}^2,m_D^2)^{3/2} }{24 \pi m_{\psi(4S)}^5}, \\
\Gamma_{\psi(4S) \to D^\ast D} &=& \frac{g_{\psi(4S)D^\ast D}^2 \lambda(m_{\psi(4S)}^2,m_{D^\ast}^2,m_D^2)^{3/2} }{6 \pi m_{\psi(4S)}^3},\label{Eq:DWDD}\\
\Gamma_{\psi(4S) \to D^\ast D^\ast} &=& \frac{g_{\psi(4S)D^\ast D^\ast}^2 \lambda(m_{\psi(4S)}^2,m_{D^\ast}^2,m_{D^\ast}^2)^{3/2} }{96 \pi m_{\psi(4S)}^5 m_{D^\ast}^4} \nonumber\\
&&\hspace{-10mm}\times (\lambda(m_{\psi(4S)}^2,m_{D^\ast}^2,m_{D^\ast}^2) +m_{\psi(4S)}^4+12m_{D^\ast}^4), \ \ \ \label{Eq:DWDStarDStar}
\end{eqnarray}
where $\lambda(x,y,z)=x^2+y^2+z^2-2xy-2xz-2yz$ denotes the K$\ddot{\mathrm{a}}$llen function. We can obtain the partial decay width related to charmed strange mesons by replacing the charmed meson masses with the corresponding masses of charmed strange mesons and multiplying a factor $1/2$ caused by isospin.

\begin{figure}[htbp]
\scalebox{0.55}{\includegraphics{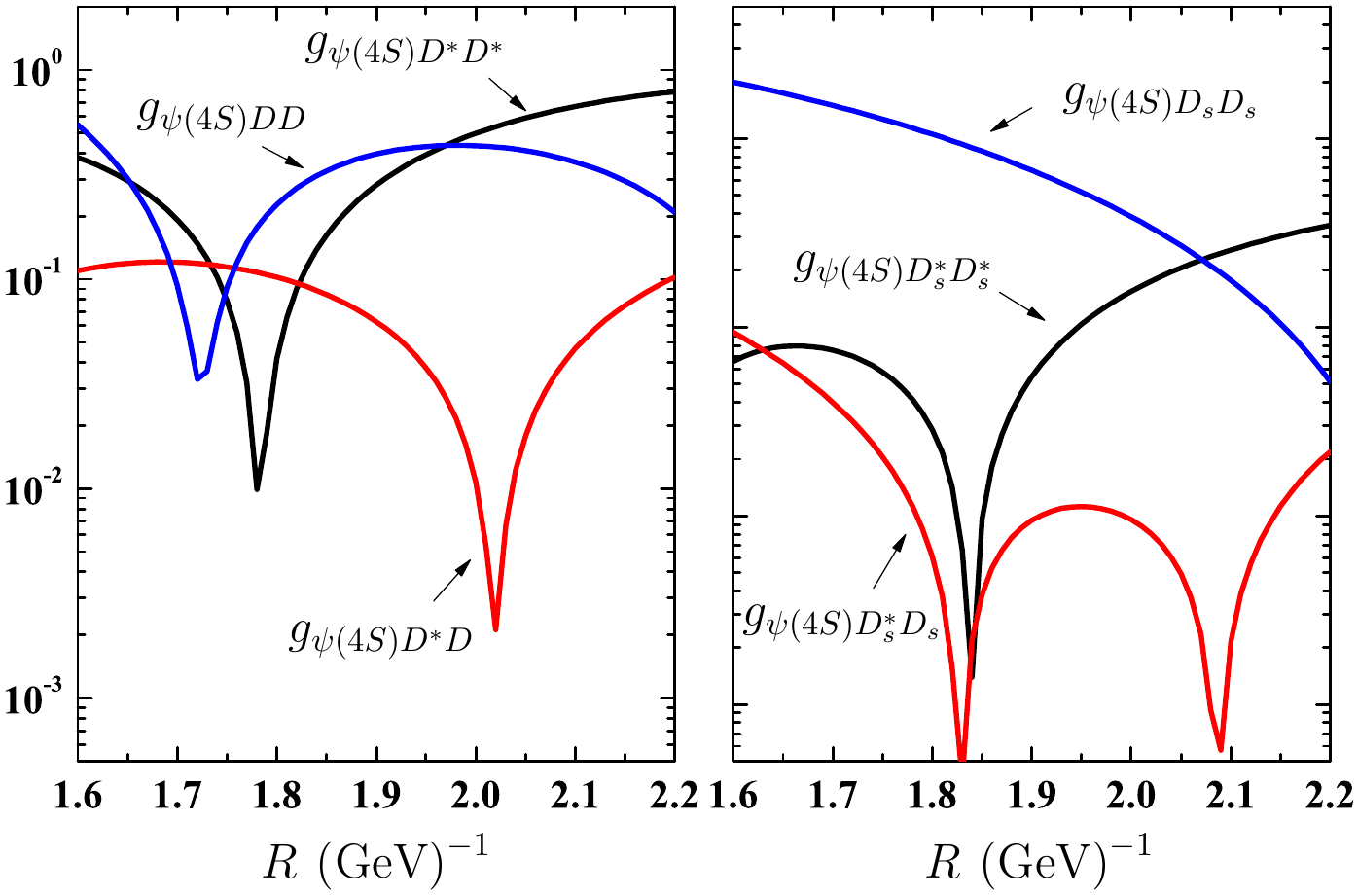}}
\caption{(color online). The $R$ dependence of the extracted coupling constants of $\psi(4S)$ interacting with charmed or charmed-strange mesons. %Here, $\psi(4S)$ is fixed to be 4230 MeV. 
\label{Fig:CP}}
\end{figure}

With the relations in Eqs. (\ref{Eq:DWDD})-(\ref{Eq:DWDStarDStar}) and the partial decay widths estimated in Ref. \cite{He:2014xna}, we can evaluate the coupling constants related to the interactions $\psi(4S)D_{(s)}^{(*)}\bar{D}_{(s)}^{(*)}$. Since the partial decay widths in Ref. \cite{He:2014xna} are dependent on the parameter $R$\footnote{{Until now, the information on the open charm decay of the predicted $\psi(4S)$ is still absent. Therefore, we adopt the theoretical results of the partial decay width of $\psi(4S)$, which were calculated by the quark pair creation (QPC) model \cite{He:2014xna}. In the QPC model, the simple harmonic oscillator (SHO) wave function is employed to describe the spatial wave functions of mesons involved in the discussed decays, where the $R$ value is the parameter in the SHO wave function
$\Psi_{n\ell m}(R,\mathbf{p})= \mathcal{R}_{n\ell}(R,\mathbf{p}) \mathcal{Y}_{\ell m}(\mathbf{p})$.}}, which is introduced in the harmonic oscillator wave function for $\psi(4S)$, the extracted coupling constants also vary with the parameter $R$, which is presented in Fig. \ref{Fig:CP}.

\renewcommand{\arraystretch}{1.4}
\begin{table}[htb]
\centering
\caption{The concrete values of coupling constants of charmonium ($J/\psi$ and $\chi_{c0}$) interacting with charmed mesons, and those of charmed mesons interacting with light pseudoscalar/vector mesons \cite{Kaymakcalan:1983qq,Oh2000qr,Casalbuoni1996pg,Colangelo2002mj}. \label{Tab:Coupling} }
\begin{tabular}{cccccc}
\toprule[1pt]
Coupling & Value &  Coupling & Value & Coupling & Value \\
\midrule[1pt]
%%%%%%%%%%%% psi
$g_{J/\psi DD}$                     & 7.44             &
$g_{J/\psi D^\ast D}$               & 2.49 GeV$^{-1}$  &
$g_{J/\psi D^\ast D^\ast}$          & 8.01             \\
%%%%%%%%%%%  Vector meson
$g_{DD\mathcal{V}}$                 & 3.47 &
$g_{D^\ast D\mathcal{V}}$           & 2.32 GeV$^{-1}$ &
$g_{D^\ast D^\ast\mathcal{V}}$      & 3.74 \\
$f_{D^\ast D^\ast\mathcal{V}}$      & 4.67 &
%%%%%%%%%%% chi_c0
$g_{\chi_{c0}DD}$                   &-25.00 GeV&
$g_{\chi_{c0}D^\ast D^\ast}$        & -8.96 GeV\\
%%%%%%%%%%% Pseudoscalar meson
$g_{D^\ast D\mathcal{P}}$           & 8.94 &
$g_{D^\ast D^\ast\mathcal{P}}$      & 17.32 GeV$^{-1}$\\
  \bottomrule[1pt]
\end{tabular}
\end{table}

Other coupling constants $\chi_{c0}D^{(\ast)} D^{(\ast)}$, $J/\psi D^{(\ast)} D^{(\ast)}$, $D^{(\ast)}D^{(\ast)} \mathcal{P}$, and $D^{(\ast)} D^{(\ast)} \mathcal{V}$ can be estimated by considering heavy quark limit and chiral symmetry, which are given in Refs. \cite{Kaymakcalan:1983qq,Oh2000qr,Casalbuoni1996pg,Colangelo2002mj}. In Table \ref{Tab:Coupling}, we list the concrete values of these coupling constants adopted in our calculation. The coupling constant involved in charmed strange mesons can be obtained by the relations $g_{Y D_s^{(\ast)} D_s^{(\ast)}} = \sqrt{m_{D_s^{(\ast)}} m_{D_s^{(\ast)}} / m_{D^{(\ast)}} m_{D^{(\ast)}}}\ g_{Y D^{(\ast)} D^{(\ast)}}$ with $Y=J/\psi,\ \chi_{c0},\ \mathcal{V},\ \mathcal{P}$ \cite{Kaymakcalan:1983qq,Oh2000qr,Casalbuoni1996pg,Colangelo2002mj}.
{In addition, the mass values of the mesons in the decays under discussion are collected in Table \ref{Tab:Mass}
\cite{Agashe:2014kda}.}

\renewcommand{\arraystretch}{1.4}
\begin{table}[htb]
\centering
\caption{The masses of the involved particles in present work. \label{Tab:Mass} }
\begin{tabular}{ccc}
\toprule[1pt]
 State&Mass  \cite{Agashe:2014kda} \\
\midrule[1pt]
$\psi(4S)$             & $4230\pm 8$  GeV \cite{Ablikim:2014qwy}\\
$J/\psi$               & $3096.9\pm 0.011$  MeV\\
$\chi_{c0}$            & $3414.75\pm0.31$  MeV \\
%%%%%%%%%%%%%%%%
$\eta$                 & $547.862\pm 0.018$ MeV\\
$\omega$               & $782.65\pm0.12$ MeV\\
$D^{0/\pm}$                    & $1864.84\pm0.07$/$1869.61\pm0.01$ MeV\\
$D^{\ast0/\pm}$               & $2006.96\pm0.10/2010.26\pm0.07$ MeV\\
$D_s$                  & $1968.3\pm0.11$ MeV\\
$D_s^\ast$             & $2112\pm0.4$ MeV\\
  \bottomrule[1pt]
\end{tabular}
\end{table}

\section{Numerical results}\label{sec3}

Before presenting the numerical results, we first focus on the BESIII measurement of $e^+e^-\to\omega\chi_{c0}$
\cite{Ablikim:2014qwy}. If explaining the observed enhancement structure existing in $e^+e^-\to\omega\chi_{c0}$ to be $\psi(4S)$, the BESIII data indicates \cite{Ablikim:2014qwy}
\begin{eqnarray}
\Gamma({\psi(4S)\to e^+e^-} )\mathcal{B}(\psi(4S) \to \omega \chi_{c0})= (2.7 \pm 0.5 \pm 0.4) \,\mathrm{eV},\nonumber\\\label{k1}
\end{eqnarray}
which can be applied to extract the branching ratio of $\psi(4S)\to \omega\chi_{c0}$. In order to get the branching ratio $\mathcal{B}(\psi(4S) \to \omega \chi_{c0})$ from Eq. (\ref{k1}), we have to rely on the theoretical evaluation of the width $\Gamma(\psi(4S)\to e^+e^-)$. In Refs. \cite{Dong:1994zj,Li:2009zu}, the screen potential was considered when studying the mass spectrum of the charmonium family. The mass of $\psi(4S)$ is calculated in Refs. \cite{Dong:1994zj,Li:2009zu} and is given by 4274 MeV and 4273 MeV, respectively, both of which are close to the mass of the enhancement structure reported by BESIII \cite{Ablikim:2014qwy}. In Refs. \cite{Dong:1994zj,Li:2009zu}, the partial width of $\psi(4S)\to e^+e^-$ was also estimated, i.e., $\Gamma(\psi(4S)\to e^+ e^-)=0.63$ keV \cite{Dong:1994zj} and $\Gamma(\psi(4S)\to e^+ e^-)=0.66$ keV \cite{Li:2009zu}. If taking theoretical range $\Gamma(\psi(4S)\to e^+e^-)=0.63 \sim 0.66$ keV, we can obtain the branching ratio $\mathcal{B}(\psi(4S)\to \omega \chi_{c0})=(3.1\sim 5.3)\times 10^{-3}$, which is the same order of the upper bound for $\mathcal{B}(\Upsilon(5S) \to \omega \chi_{b0})$ \cite{He:2014sqj}. This extracted $\mathcal{B}(\psi(4S)\to \omega \chi_{c0})$ ratio will be compared with our calculation.

\begin{figure}[htbp]
\scalebox{0.45}{\includegraphics{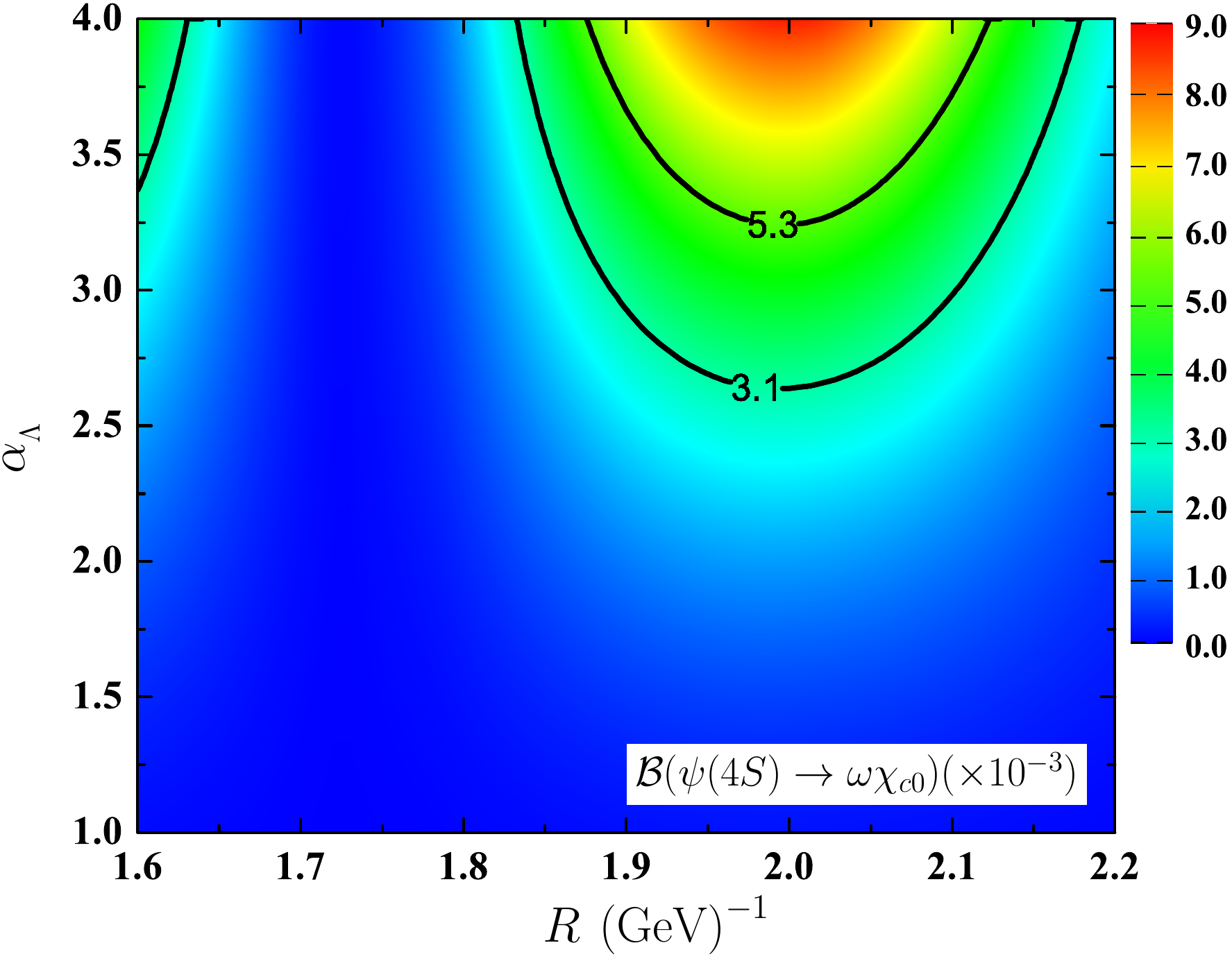}}
\caption{(color online). The branching ratio of $\psi(4S) \to \omega \chi_{c0}$ depending on the $R$ value and parameter $\alpha_{\Lambda}$. Comparison with the extracted experimental data is shown in black solid curves \cite{Ablikim:2014qwy}. \label{Fig:OmegaChic0}}
\end{figure}

Using the formula given above, we calculate the ${\psi(4S) \to \omega \chi_{c0}}$. In Fig. \ref{Fig:OmegaChic0}, we present the branching ratio of ${\psi(4S) \to \omega \chi_{c0}}$ dependent on $\alpha_{\Lambda}$ and $R$ values, where we use the theoretical total decay width of $\psi(4S)$ to transfer the obtained partial decay width to the branching ratio \cite{He:2014xna}. {Actually, the calculated total and partial widths of the predicted $\psi(4S)$ in Ref. \cite{He:2014xna} are model-dependent. For example, the total and partial widths are proportional to the square of the strength of quark-antiquark pair creation, i.e., $\Gamma_{\mathrm{total}} \propto \gamma^2$ and $\Gamma_{i} \propto \gamma^2$  \cite{He:2014xna}, where $\Gamma_{\mathrm{total}}$ and $\Gamma_{i}$ denote the total and partial widths, respectively. Although the uncertainty of $\gamma$ leads to the uncertainties of the total and partial decay widths,
there does not exist propagation of the uncertainty of $\gamma$ on the calculated branching ratios of $\psi(4S)\to \omega\chi_{c0}$ and $\psi(4S)\to \eta J/\psi$, since these calculated branching ratios are independent on the $\gamma$ value.} The upper and lower limits of branching ratio obtained from the experimental measurement are also given in Fig. \ref{Fig:OmegaChic0}. From our present calculation, we find that our theoretical result overlaps with the experimental data in a reasonable parameter range, {which is sandwiched by two curves in Fig. \ref{Fig:OmegaChic0}}. This study shows that the $e^+e^-\to \omega\chi_{c0}$ observation can be understood through introduction of the predicted $\psi(4S)$ contribution, {which implies that it is reasonable to identify the enhancement near 4.2 GeV in the $e^+e^- \to \omega \chi_{c0}$ cross section as the missing $\psi(4S)$}.

\iffalse
\begin{table}[htb]
\centering
\caption{The mass and dilepton decay width of $\psi(nS)$ evaluated by quark model comparing with the experimental data \cite{Agashe:2014kda}. The masses and dilepton decay widths are in unit of MeV and keV, respectively.\label{Tab:dilepton}}
\begin{tabular}{c|cc|cc|cc}
\hline\hline
  \multirow{2}*{State}  &\multicolumn{2}{c|}{ Expt.}    &\multicolumn{2}{c|}{ Ref. \cite{Dong:1994zj}}    &  \multicolumn{2}{c}{Ref. \cite{Li:2009zu}}\\
\cline{2-7}
    & Mass & $\Gamma_{ee}$  &Mass  & $\Gamma_{ee}$  &Mass  & $\Gamma_{ee}$\\
\hline
$\psi(1S)$ &  $3097$  & $5.55 \pm 0.14 \pm 0.02$  & 3.097  & 4.34 &  3097 & 4.54\\
$\psi(2S)$ &  $3686$  & $2.35 \pm 0.04 $          & 3643   & 1.90 &  3673 & 1.66\\
$\psi(3S)$ &  $4039$  & $0.86 \pm 0.07 $          & 4000   & 1.09 &  4023 & 0.98\\
$\psi(4S)$ &  $4230$  & ---                       & 4274   & 0.63 &  4273 & 0.66 \\
\hline
\end{tabular}
\end{table}
\fi

\begin{figure}[htbp]
\scalebox{0.45}{\includegraphics{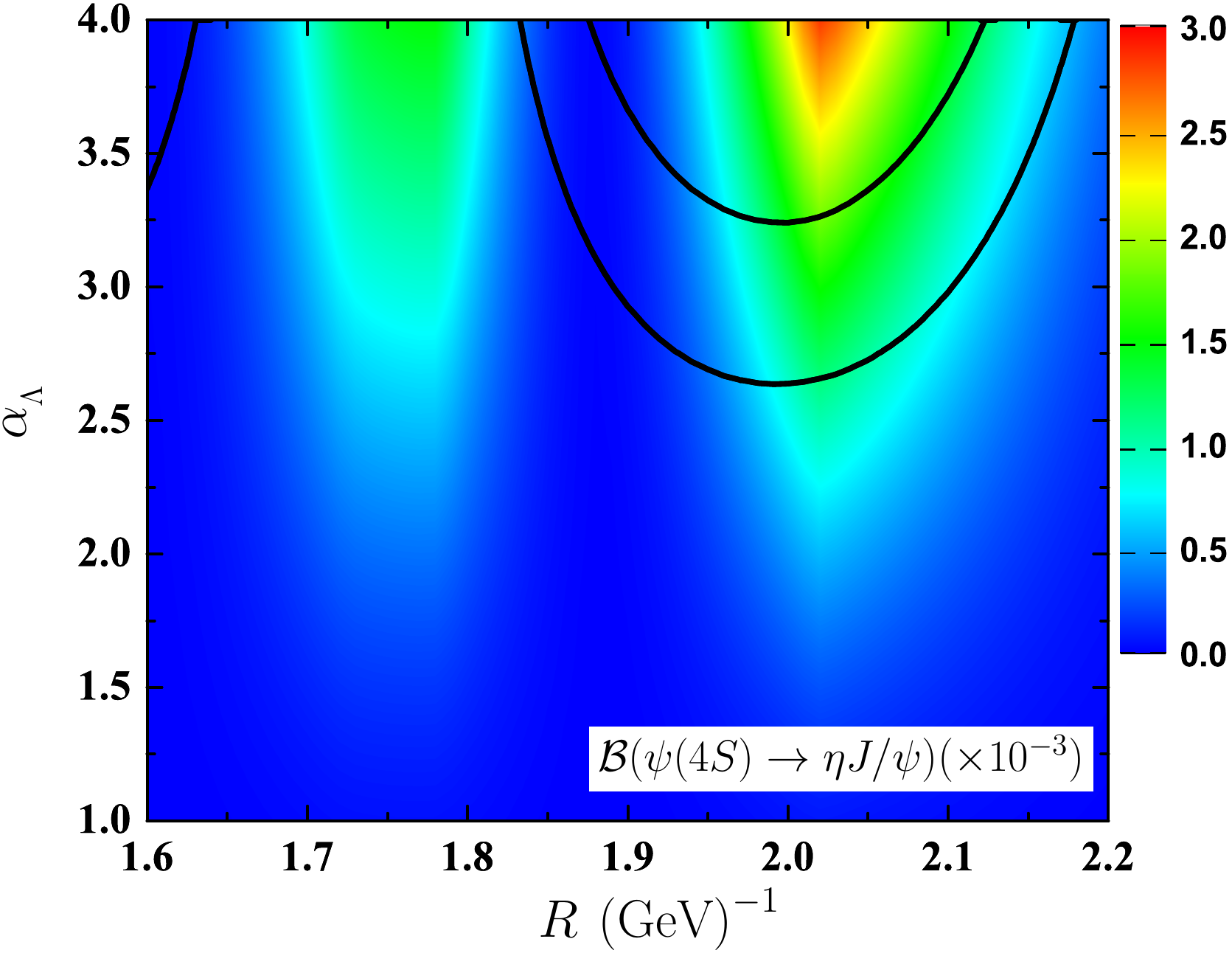}}
\caption{(Color online) The branching ratio of $\psi(4S) \to \eta J/\psi$ dependent on the $R$ value and parameter $\alpha_{\Lambda}$. Here, the range sandwiched by the black curves gives a possible branching ratio of $\psi(4S) \to \eta J/\psi$ corresponding to parameter range, which are the constraints given by the study of $\psi(4S) \to \omega \chi_{c0}$.
\label{Fig:EtaJpsi}}
\end{figure}

In the same way as $\psi(4S) \to \omega \chi_{c0}$, we can also study the hidden charm decay process $\psi(4S) \to \eta J/\psi$ via the hadronic loop mechanism\footnote{In Ref. \cite{Chen:2012nva}, we have estimated the meson loop contributions to the $\eta$ transition between $\psi(4040)/\psi(4160)$ and $J/\psi$, which are consistent with the experimental measurements.}. The phase space of $\psi(4S) \to J/\psi \eta $ is larger than that of $\psi(4S)\to \omega \chi_{c0}$. $\psi(4S)\to \eta J/\psi$ occurs via $P$-wave while $\psi(4S)$ decays into $\omega \chi_{c0}$ through $S$-wave, which is the difference between $\psi(4S)\to \eta J/\psi$ and $\psi(4S)\to\omega \chi_{c0}$.

In Fig. \ref{Fig:EtaJpsi}, the $R$ and $\alpha_\Lambda$ dependence of the branching ratio $\psi(4S)\to \eta J/\psi$ is presented. If taking the same parameter range as that of  $\psi(4S)\to\omega \chi_{c0}$, we find that there is large variation between the upper and lower limits of the branching ratio of $\psi(4S)\to \eta J/\psi$. Thus, in this work we incline to predict the upper limit of the branching ratio, i.e.,  $$\mathcal{B}(\psi(4S)\to \eta J/\psi)<1.9 \times 10^{-3},$$ which can be tested in future experiments, especially BESIII and forthcoming BelleII. % {\color{blue} At present, the data of the cross section of $e^+ e^- \to J/\psi \eta$ are available for Belle Collaboration \cite{Wang:2012bgc}, we suggest the Belle Collaboration to carry out the fit including $\psi(4040)$, $\psi(4160)$ and $\psi(4S)$ under current statistics}

{Calculation of a decay width has a close relationship with input parameters like coupling constants and masses. Since the involved coupling constants and masses carry error bars, we should further discuss error contributions to final results. However, we notice that most of the coupling constants listed in Table \ref{Tab:Coupling} were given in Refs. \cite{Kaymakcalan:1983qq,Oh2000qr,Casalbuoni1996pg,Colangelo2002mj} without including their errors. Thus, in this work it is difficult to discuss error effects of coupling constants to the decay widths we calculate. In Table \ref{Tab:Mass}, we show the masses of the involved mesons with errors, which show that measurement of their masses is very precise since their masses carry very small errors except for $\psi(4S)$. In the following, we only discuss the dependence of decay widths on the error of the $\psi(4S)$ mass.}

\begin{figure}[htpb]
\scalebox{0.55}{\includegraphics{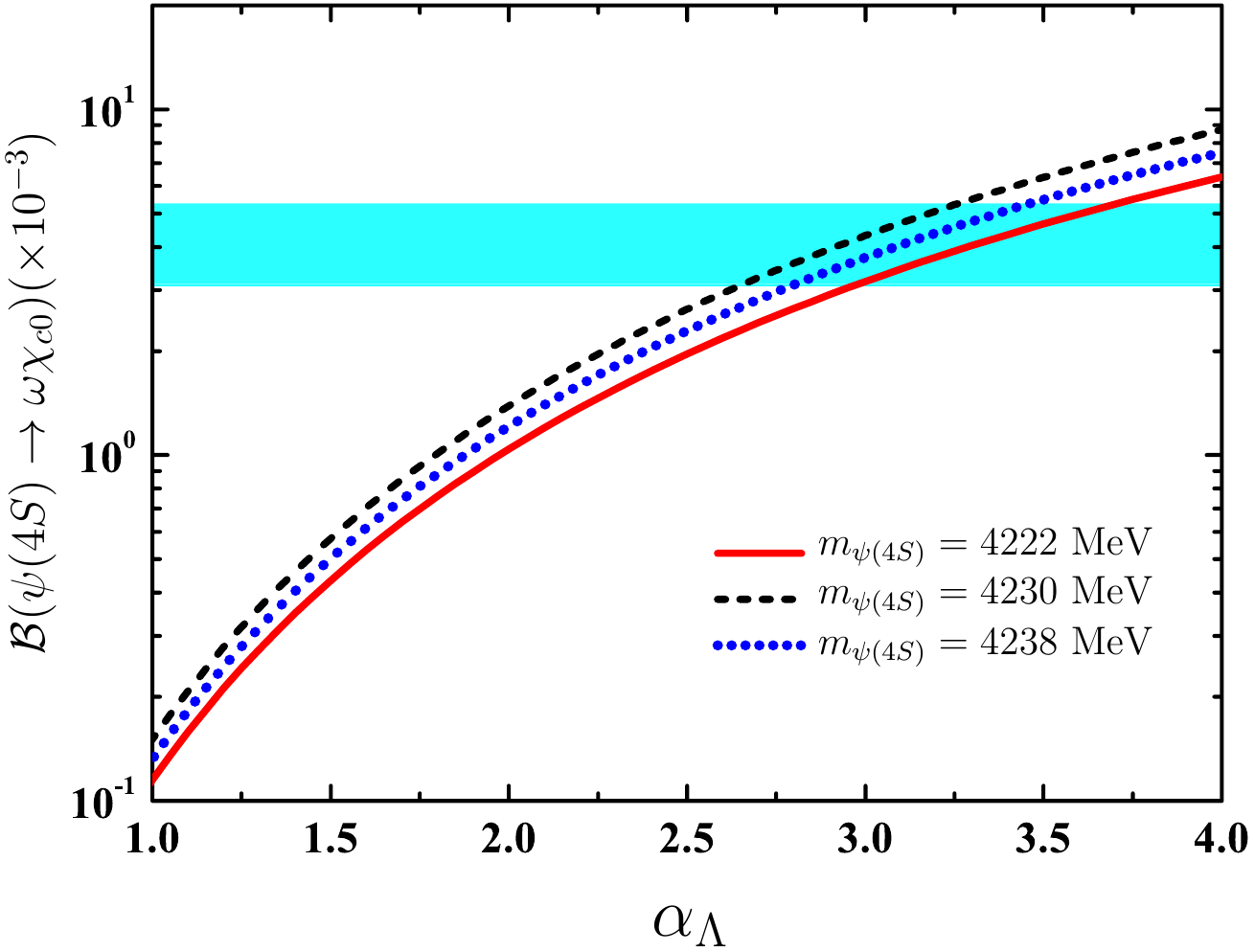}}
\caption{(Color online). The $\alpha_{\Lambda}$ dependence of $\mathcal{B}(\psi(4S) \to \omega \chi_{c0})$ with different values of the $\psi(4S)$ mass. The horizontal band is the extracted experimental data, which correspond to $3.1 \times 10^{-3} < \mathcal{B}(\psi(4S) \to \omega \chi_{c0}) < 5.3 \times 10^{-3}$.  \label{Fig:Massdep}}
\end{figure}

{If considering the error bar of the $\psi(4S)$ mass, the partial and total decay widths of $\psi(4S)$ will depend on both the mass of $\psi(4S)$ and the $R$ value. To discuss the mass dependence of the branching ratio of $\psi(4S) \to \omega \chi_{c0}$, we take a typical value of $R=2.0\ \mathrm{GeV}^{-1}$. In Fig. \ref{Fig:Massdep}, we present the $\alpha_{\Lambda}$ dependence of the branching ratio of $\psi(4S) \to \omega \chi_{c0}$ with three values of the $\psi(4S)$ mass, which are 4222 MeV, 4230 MeV, and 4238 MeV, respectively, and are within an error range. From the figure, one can conclude that the $\psi(4S)$ mass weakly affects the branching ratio of $\psi(4S) \to \omega \chi_{c0}$.}

%we give the band of the coupling constants of interactions $\psi(4S)D_{(s)}^{(*)}\bar D_{(s)}^{(*)}$ (see Fig. \ref{Fig:CP:error}). Since our calculation is related two free parameters $R$ and $\alpha_\Lambda$, thus we take typical $R$ and $\alpha_\Lambda$ values to discuss the above issue.}

%{\color{red}With $\psi(4S)\to \chi_{c0}\omega$ as an example, }

\section{Conlcusions and discussion}\label{sec4}

In this work, we have proposed that the newly observed $e^+e^-\to \omega\chi_{c0}$ by BESIII \cite{Aubert:2005rm} can be due to the contribution from the missing charmonium $\psi(4S)$, which was predicted in Ref. \cite{He:2014xna}
by the similarity between charmonium and bottomonium families. This proposal is supported by the comparison between the resonance parameter of the reported structure in $e^+e^-\to \omega\chi_{c0}$ \cite{Aubert:2005rm}, the theoretical results in Ref. \cite{He:2014xna}, and the estimated mass of $\psi(4S)$ via the screen potential \cite{Dong:1994zj,Li:2009zu}.

If the above assumption is correct, we must understand $e^+e^-\to \omega\chi_{c0}$ process by studying the $\psi(4S)\to \omega\chi_{c0}$ decay and comparing this calculation with the corresponding extracted experimental data.
Accordingly, in the present work we have studied the $\psi(4S)\to \omega\chi_{c0}$ decay mediated by the hadronic loop mechanism.
Our theoretical calculation has shown that the extracted branching ratio of the $\psi(4S)\to \omega\chi_{c0}$ transition can be well described, which provides a direct support to our proposal.

{We need to specify that there exists $\alpha_\Lambda$ dependence of the calculated branching ratio of $\psi(4S)\to \omega\chi_{c0}$\footnote{We thank the anonymous referee for pointing out this fact.}. At present, the definite value of $\alpha_\Lambda$ is unknown and cannot be fixed by other relevant processes, which makes us difficult to give a definite conclusion that the BESIII observation of $e^+e^-\to \chi_{c0}\omega$ is from the $\psi(4S)$ contribution. In this work, we only give a possible explanation to the process $e^+e^-\to \chi_{c0}\omega$ experimentally observed by introducing the $\psi(4S)$ contribution since the extracted branching ratio of $\psi(4S)\to \omega\chi_{c0}$ can be reproduced with some typical and reasonable $\alpha_\Lambda$ values. More theoretical and experimental joint efforts will be helpful to clarify this point.
}

As a prediction, the upper limit of the branching ratio of
$\psi(4S)\to \eta J/\psi$ has been given, which is similar to the discussed $\psi(4S)\to\omega \chi_{c0}$. The predicted upper limit of $\psi(4S)\to \eta J/\psi$ indicates that $\psi(4S)\to \eta J/\psi$ can be accessible at future experiment, especially at BESIII, Bellle and forthcoming BelleII, where we have also suggested a measurement of $e^+e^-\to \eta J/\psi$ to be carried out.
In Ref. \cite{Wang:2012bgc}, the measurement of the $J/\psi\eta$ in variant mass distribution of $e^+e^-\to J/\psi\eta$ was given by Belle. We notice that there exists a narrow structure around 4.23 GeV. Thus, we suggest Belle to carry out a further fit through including $\psi(4040)$, $\psi(4160)$, and the predicted $\psi(4S)$ with narrow width.

Before closing this work, we should mention the measurement of the cross section of $e^+ e^- \to \pi^+ \pi^- h_c$ at $\sqrt{s}=3.90 \sim4.42$ GeV done by BESIII \cite{Ablikim:2013wzq}. Here, the measured cross section of $e^+ e^- \to \pi^+ \pi^- h_c$ is of the same order of magnitude as that of the $e^+e^-\to \pi^+ \pi^- J/\psi$. However, the lineshape of $e^+e^- \to \pi^+ \pi^- h_c$ is different from that of $e^+e^-\to \pi^+ \pi^- J/\psi$. By fitting the available experimental data of $e^+ e^- \to \pi^+ \pi^- h_c$ from 3.90 to 4.42 GeV, a narrow structure around 4.2 GeV was discovered, where the mass and width are reported to be $M=4216 \pm 7 $ MeV and $\Gamma=39 \pm 17$ MeV or $M=4230 \pm 10$ MeV and $\Gamma=12 \pm 36$ MeV \cite{Yuan:2013ffw}, which depend on the different assumptions of lineshape trend above 4.42 GeV. In Ref. \cite{He:2014xna}, the authors have once suggested that this narrow structure existing in $e^+ e^- \to \pi^+ \pi^- h_c$ can be due to the predicted missing $\psi(4S)$.

\begin{table}[htpb]
\centering
\caption{Summary of resonance parameters of the structures reported in the $e^+e^-\to\omega\chi_{c0}$ and $e^+e^-\to\pi^+ \pi^-h_c$ processes.\label{Tab:Psi4S_Exp} }
\begin{tabular}{cccc}
\toprule[1pt]
Process  &  Mass (MeV)  &  Width (MeV)  \\
\midrule[1pt]
$e^+e^-\to\omega \chi_{c0}$ \cite{Aubert:2005rm}& $4230 \pm 8 \pm 6$ & $ 38 \pm 12 \pm2$   \\
\multirow{2}*{$e^+e^-\to\pi^+ \pi^- h_c$ \cite{Yuan:2013ffw}} & $4216 \pm 7$ & $39 \pm 17$ \\
                                 & $4230 \pm 10$ & $12 \pm 36$ \\
  \bottomrule[1pt]
\end{tabular}
\end{table}

We notice that the cross sections of the $e^+ e^-$ annihilation into a pair of charmed and/or charmed-strange mesons have been measured in Refs. \cite{Aubert:2009aq, Pakhlova:2009jv, Pakhlova:2007fq, Abe:2006fj, Pakhlova:2008zza} to search for charmonium-like state $Y(4260)$. The analysis presented in Refs. \cite{Aubert:2009aq, Pakhlova:2009jv, Pakhlova:2007fq, Abe:2006fj, Pakhlova:2008zza} shows no evident signal of $Y(4260)$ in its open-charm decay channels.
This is because $Y(4260)$ is totally different from the predicted $\psi(4S)$ since the widths of $Y(4260)$ and predicted $\psi(4S)$ are quite different from each other, i.e., $Y(4260)$ is a broad state with the width $120\pm12$ MeV \cite{Agashe:2014kda}, while the predicted $\psi(4S)$ have a narrow width as indicated in Ref. \cite{He:2014xna}, which can be the reason why the predicted $\psi(4S)$ is not described in the analysis of the open-charm decay channels \cite{Aubert:2009aq, Pakhlova:2009jv, Pakhlova:2007fq, Abe:2006fj, Pakhlova:2008zza} and the $R$ value scan \cite{Burmester:1976mn,Brandelik:1978ei,Siegrist:1981zp,Osterheld:1986hw,Bai:1999pk,Bai:2001ct,CroninHennessy:2008yi,Ablikim:2009ad}. 

{For the predicted $\psi(4S)$, it has open-charm decay channels \cite{He:2014xna}, which means that there should be a $\psi(4S)$ evidence in the experimental data of the $e^+e^-\to D_{(s)}^{(*)}\bar{D}_{(s)}^{(*)}$ processes. When carefully checking the Belle  \cite{Pakhlova:2008zza} and BaBar \cite{Aubert:2009aq} data of of $e^+e^- \to D\bar{D}$ , we find that there exists a narrow structure around 4.2 GeV, which can correspond to the missing $\psi(4S)$ we have predicted in the present work (see Fig. \ref{Fig:DD} for more details). We need to specify that Belle and BaBar \cite{Pakhlova:2008zza,Aubert:2009aq} adopted the same bin size (20 MeV) in their analysis. If comparing the Belle result with the BaBar data  just shown in Fig. \ref{Fig:DD}, we still find that there exists obvious difference between the Belle  \cite{Pakhlova:2008zza} and BaBar  \cite{Aubert:2009aq} data. For example, the concrete position of the peak around 4.2 GeV shown in the BaBar data is lower than that given in the Belle data. This experimental situation of $e^+e^-\to D_{(s)}^{(*)}\bar{D}_{(s)}^{(*)}$ should be clarified by further experimental effort. 

\begin{figure}[htbp]
\scalebox{0.6}{\includegraphics{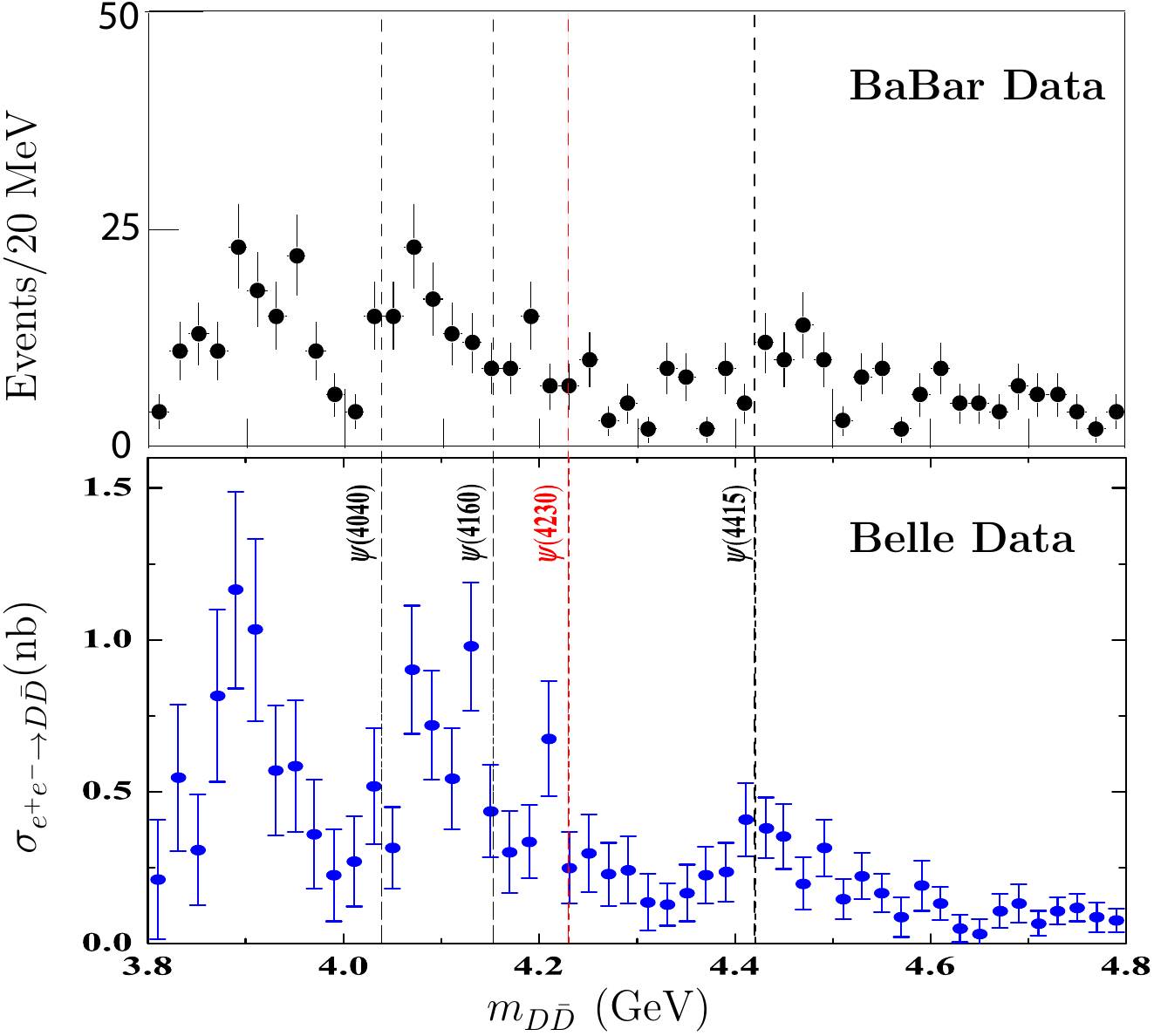}}
\caption{(Color online) The experimental data of $e^+e^- \to D\bar{D}$ from the Belle \cite{Pakhlova:2008zza} and BaBar \cite{Aubert:2009aq} collaborations and the comparison with the central masses of $\psi(4040)$, $\psi(4160)$, $\psi(4415)$ and the predicted $\psi(4S)$. Here, the predicted mass is taken from Ref. \cite{Aubert:2005rm}.
\label{Fig:DD}}
\end{figure}

Besides the experimental data of $e^+e^-\to D\bar{D}$ \cite{Pakhlova:2008zza,Aubert:2009aq}, experiment also reported other measurements of the open-charm decay channels from the $e^+e^-$ annihilation \cite{Pakhlova:2009jv,Pakhlova:2007fq,Abe:2006fj,Aubert:2009aq}, i.e.,  the cross sections of $e^+e^-\to D^{*+}D^{*-}$ and $e^+e^-\to D^{+}D^{*-}$ were reported by Belle in \cite{Abe:2006fj}, while the event of $e^+e^-\to D^*\bar{D}$ and $e^+e^-\to D^*\bar{D}^*$ dependent on 
the invariant mass distributions $M(D^*\bar{D})$ and  $M(D^*\bar{D}^*)$, respectively, was given by BaBar \cite{Aubert:2009aq}, where both Belle and BaBar adopted 20 MeV bin size in the analysis. In addition, Belle also measured the cross sections of $e^+e^-\to D^0D^{-}\pi^+$ \cite{Pakhlova:2007fq} and $e^+e^-\to D^0D^{*-}\pi^+$ \cite{Pakhlova:2009jv} dependent on the mass distributions $M(D^0D^{-}\pi^+)$ and $M(D^0D^{*-}\pi^+)$, respectively. If checking the analysis in Refs. \cite{Pakhlova:2009jv,Pakhlova:2007fq}, we find that the bin size is about 40 MeV, which is larger than the width of the predicted $\psi(4S)$. It is difficult to identify the signal of the predicted $\psi(4S)$ by the measured cross sections in Refs. \cite{Pakhlova:2009jv,Pakhlova:2007fq}.

However, we also notice the experimental data released in Refs. \cite{Abe:2006fj,Aubert:2009aq}, where the bin size is taken as 20 MeV and one data collected per bin. There are two data in the region of the resonance for the $D\bar{D}^*$, and $D^*\bar{D}^*$ channels \cite{Abe:2006fj,Aubert:2009aq}, but those are compatible with the
background and does not reflect any resonance\footnote{We thank the anonymous referee for reminding us this information.}. Just shown in Table \ref{Tab:Psi4S_Exp}, the measurements of the width of the structures existing in the $e^+e^-\to \omega\chi_{c0}$ \cite{Aubert:2005rm} and $e^+e^-\to \pi^+\pi^- h_c$ \cite{Yuan:2013ffw} processes are different from each other. If considering experimental $12\pm36$ MeV \cite{Yuan:2013ffw}, where this central value of width is smaller than the bin size adopted in Refs. \cite{Abe:2006fj,Aubert:2009aq}, it is still difficult to 
observe a corresponding enhancement through the data of the open-charm decay channels listed in Refs. \cite{Abe:2006fj,Aubert:2009aq}.
In addition, the experimental fact shown in Refs. \cite{Abe:2006fj,Aubert:2009aq} can also provide an extra support to the peculiarity of the predicted $\psi(4S)$, i.e., $\psi(4S)$ should be a very narrow state with full width smaller than 20 MeV \cite{He:2014xna}, since there does not exist any enhancement around 4230 MeV by analyzing the data with 20 MeV bin size of the $D\bar{D}^*$, and $D^*\bar{D}^*$ channels \cite{Abe:2006fj,Aubert:2009aq}.

Thus, the resonance parameter of $\psi(4S)$ is a crucial information. Considering this status, we firstly suggest more precise measurement of the resonance parameters of the narrow structures in $e^+e^-\to \omega\chi_{c0}$ \cite{Aubert:2005rm} and $e^+e^-\to \pi^+\pi^- h_c$ \cite{Yuan:2013ffw}. The further experimental result of the $D\bar{D}$, $D\bar{D}^*$, and $D^*\bar{D}^*$ channels can be applied to test this $\psi(4S)$ assignment proposed in the present work. }

Secondly, we still need more experimental data of 
the open-charm decays from the $e^+e^-$ annihilation and $R$ value scan, where the small bin size should be adopted. These experimental efforts will be useful to give a definite conclusion of whether the predicted $\psi(4S)$ can be found in these channels. BESIII and forthcoming BelleII have good opportunity to carry out such precise measurements in the following years.

At present, we can find two experimental evidences for existence of a narrow charmonium $\psi(4S)$, where we collect the experimental information of the narrow structures near 4.2 GeV in Table \ref{Tab:Psi4S_Exp}. Comparing the experimental data listed in Table \ref{Tab:Psi4S_Exp}, we notice that these data are comparable with each other when considering the experimental errors. Combining them with our prediction of the branching ratio of $\psi(4S)\to \eta J/\psi$, we expect that there exists a similar narrow structure near 4.2 GeV in the $\eta J/\psi$ distribution of $e^+e^-\to \eta J/\psi$.
{In Ref. \cite{Wang:2012bgc}, Belle measured the $\eta J/\psi$ invariant mass distribution of $e^+e^-\to \eta J/\psi$.
Thus, we suggest Belle to redo the analysis by including the predicted $\psi(4S)$, which is an interesting issue. We also expect more experimental progresses in future experiments, especially Belle, BESIII, and forthcoming BelleII.

%\vfil

\noindent {\bf Acknowledgments}:
This work is supported by the National Natural Science Foundation
of China under Grants No. 11222547, No. 11375240, No. 11175073, and
No. 11035006, and the Ministry of Education of China (SRFDP under Grant
No. 2012021111000), and the Fok Ying Tung Education Foundation (No. 131006).

%\iffalse
\noindent{\bf Note added}: After submitting our paper to arXiv, we have noticed a very recent work of $e^+e^-\to \omega \chi_{c0}$ in arXiv \cite{Li:2014jja}, in which the authors suggested that  $e^+e^-\to \omega \chi_{c0}$ can be due to the intermediate $\psi(4160)$ contribution.

\section*{Appendix}
The Feynman rules corresponding to effective Lagrangians listed in Eq. (\ref{Eq:LpsiDD}) are
%%
%%  psi DD
%%
\begin{eqnarray}
\raisebox{-15pt}{\includegraphics[width=0.15%
\textwidth]{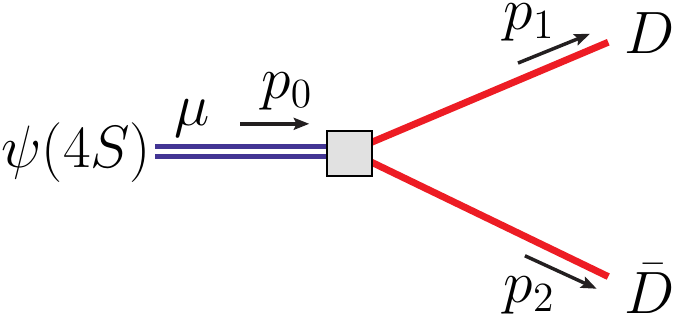}} &\widehat{=} & ig_{\psi \mathcal{D} \mathcal{D}} (ip_{2\mu} -ip_{1\mu}),\label{Eq:VpsiDD}\\
%\end{eqnarray}
%%
%%  psi DStar Dbar
%%
%\begin{eqnarray}
\raisebox{-15pt}{\includegraphics[width=0.15%
\textwidth]{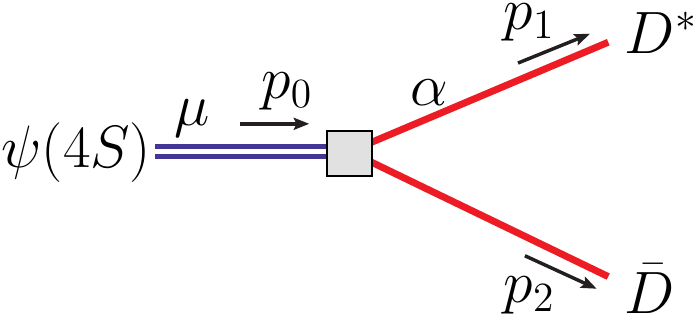}} &\widehat{=}& g_{\psi \mathcal{D}^\ast \mathcal{D}} \varepsilon_{\theta \mu \alpha \tau} (-ip_0^\theta) (ip_2^\tau-ip_1^\tau),\\
%\end{eqnarray}
%%
%%  psi D  DStarbar
%%
%\begin{eqnarray}
\raisebox{-15pt}{\includegraphics[width=0.15%
\textwidth]{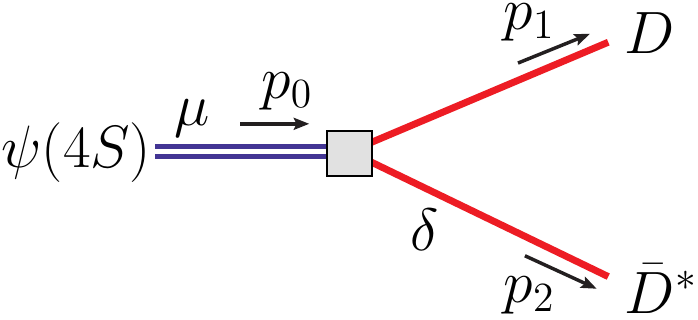}} &\widehat{=}&  g_{\psi \mathcal{D}^\ast \mathcal{D}} \varepsilon_{\theta \mu \alpha \tau} (-ip_0^\theta) (i p_1^\tau-ip_2^\tau),\ \ \\
%\end{eqnarray}
%%
%%  psi DStar  DStarbar
%%
%\begin{eqnarray}
\raisebox{-15pt}{\includegraphics[width=0.15%
\textwidth]{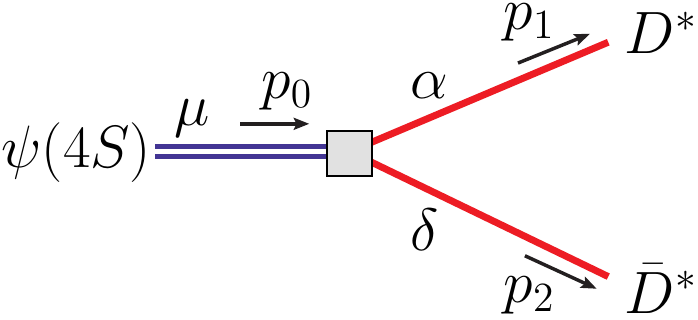}} &\widehat{=}&  i g_{\psi \mathcal{D}^\ast \mathcal{D}^\ast} (ip_{2}^{\alpha} g_\mu^\delta -ip_1^\delta g_\mu^\alpha \nonumber\\&&-i(p_{2\mu} -p_{1\mu})g_{\alpha \delta}).\label{Eq:VpsiDStarDStar}
\end{eqnarray}

%\begin{minipage}{0.15\textwidth}
%\includegraphics[width=\textwidth]{G1D.pdf}
%\end{minipage}
%\begin{minipage}{0.31\textwidth}
%\begin{eqnarray}
%&\widehat{=}&  i g_{\psi \mathcal{D}^\ast \mathcal{D}^\ast} \epsilon_{\psi}^{\mu} (ip_{2}^{\alpha} g_\mu^\delta -ip_1^\delta %g_\mu^\alpha \nonumber\\&&-i(p_{2\mu} -p_{1\mu})g_{\alpha \delta}).
%\end{eqnarray}
%\end{minipage}

With Eq. (\ref{Eq:Lchic0DD}), we can obtain the Feynman rules related to the $\chi_{c0} D^{(\ast)} D^{(\ast)}$ couplings, i.e.,
\begin{eqnarray}
\raisebox{-25pt}{\includegraphics[width=0.15%
\textwidth]{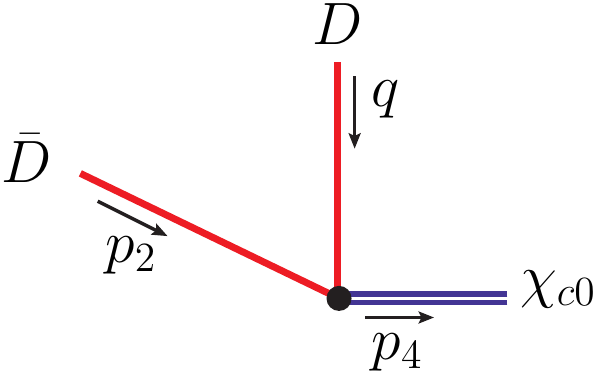}} &\widehat{=}& -g_{\chi_{c0} \mathcal{D} \mathcal{D}},\\
%\end{eqnarray}
%%
%%
%%
%\begin{eqnarray}
\raisebox{-25pt}{\includegraphics[width=0.15%
\textwidth]{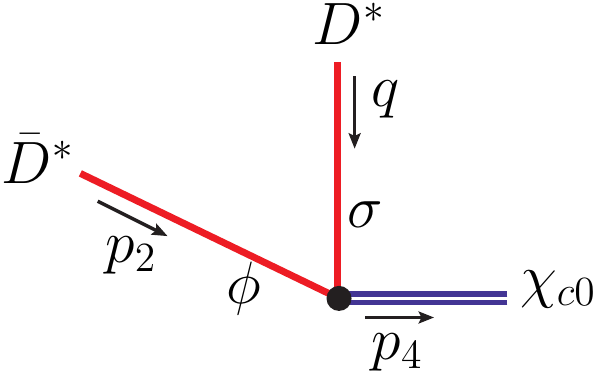}} &\widehat{=}& -g_{\chi_{c0} \mathcal{D}^\ast \mathcal{D}^\ast} g^{\phi \sigma}.
\end{eqnarray}
Next, we get the Feynman rules from Eq. (\ref{Eq:LDDV}), which depicts the coupling of charmed mesons with $\omega$ meson,
\begin{eqnarray}
\raisebox{-25pt}{\includegraphics[width=0.13%
\textwidth]{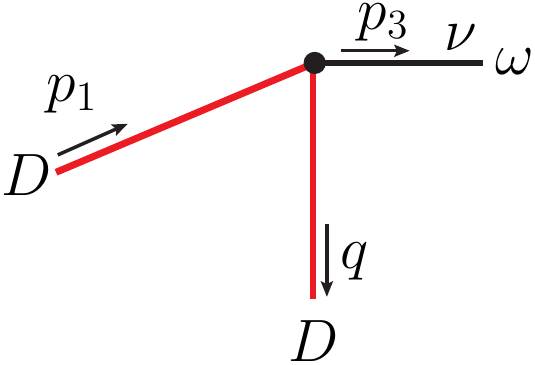}} &\widehat{=}& -ig_{\mathcal{D} \mathcal{D} \mathcal{V}} (-i(p_{1 \nu} +q_\nu )) \epsilon_{\mathcal{V}}^\nu ,\\
%\end{eqnarray}
%%
%%
%%
%\begin{eqnarray}
\raisebox{-25pt}{\includegraphics[width=0.13%
\textwidth]{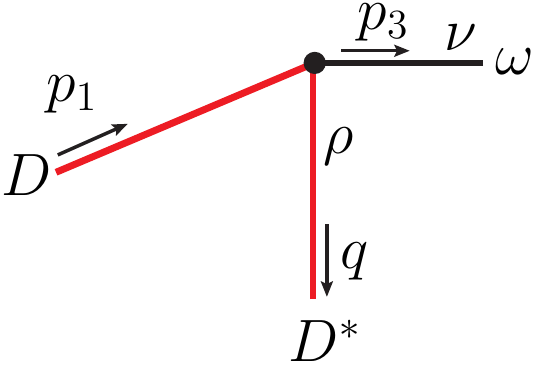}} &\widehat{=}& -2 f_{\mathcal{D}^\ast \mathcal{D} \mathcal{V}} \varepsilon_{\theta \nu \tau \rho} (ip_{3}^\theta) \epsilon_{\mathcal{V}}^\nu (ip_1^\tau+q^\tau) ,\hspace{8mm}\\
%\end{eqnarray}
%%
%%
%%
%\begin{eqnarray}
\raisebox{-25pt}{\includegraphics[width=0.13%
\textwidth]{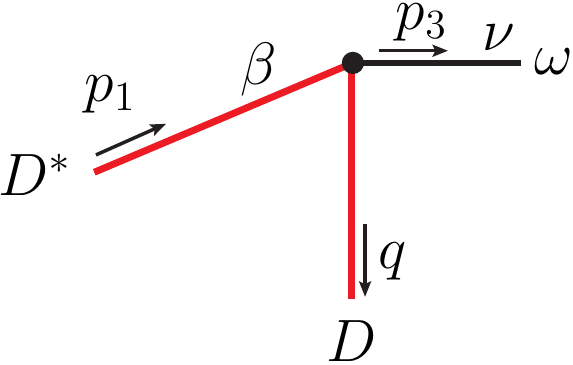}} &\widehat{=}& -2 f_{\mathcal{D}^\ast \mathcal{D} \mathcal{V}} \varepsilon_{\theta \nu \tau \rho} (ip_{3}^\theta) \epsilon_{\mathcal{V}}^\nu (-ip_1^\tau-q^\tau) ,\\
%\end{eqnarray}
%%
%%
%%
%\begin{eqnarray}
\raisebox{-25pt}{\includegraphics[width=0.13%
\textwidth]{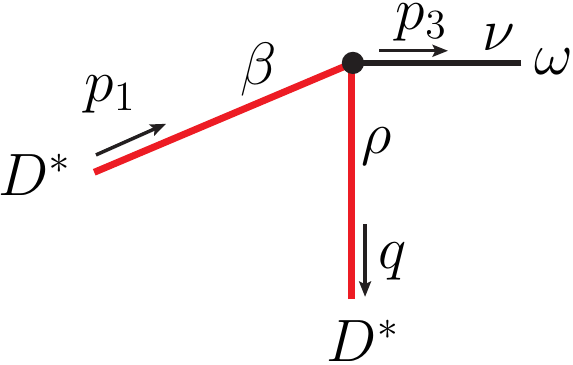}} &\widehat{=}& i g_{\mathcal{D}^\ast \mathcal{D}^\ast \mathcal{V}}  \epsilon_{\mathcal{V}}^\nu (-ip_{1\nu} -iq_\nu) g^{\beta \rho} \nonumber\\ &&+4if_{\mathcal{D}^\ast \mathcal{D}^\ast \mathcal{V}} \epsilon_{\mathcal{V}}^\nu (ip_3^\rho g^\beta_\nu -ip_3^\beta g^\rho_\nu).
\end{eqnarray}
Similarly, the Feynman rules corresponding to the interaction of charmed mesons with $\eta$ meson can be easily extracted from Eq. (\ref{Eq:LDDP}), i.e.,
%%
%% D DStar P
%%
\begin{eqnarray}
\raisebox{-25pt}{\includegraphics[width=0.13%
\textwidth]{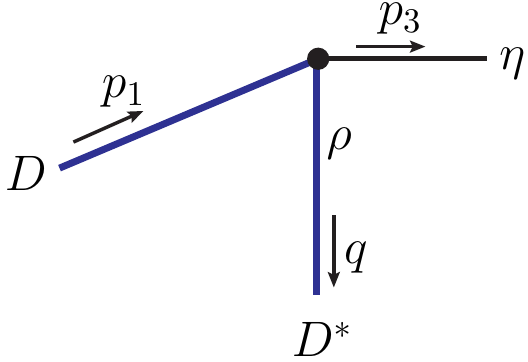}} &\widehat{=}& -ig_{\mathcal{D}^\ast \mathcal{D} \mathcal{P}} (-ip_3^\rho) ,\\
\raisebox{-25pt}{\includegraphics[width=0.13%
\textwidth]{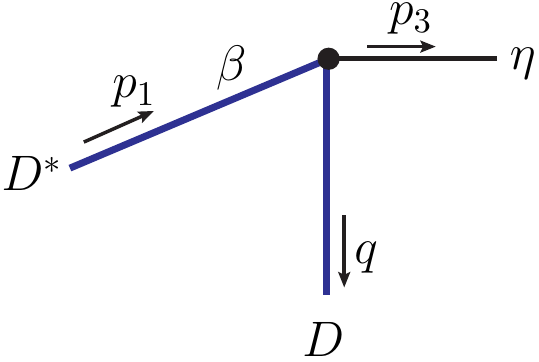}} &\widehat{=}& -ig_{\mathcal{D}^\ast \mathcal{D} \mathcal{P}} (ip_3^\rho),\\
\raisebox{-25pt}{\includegraphics[width=0.13%
\textwidth]{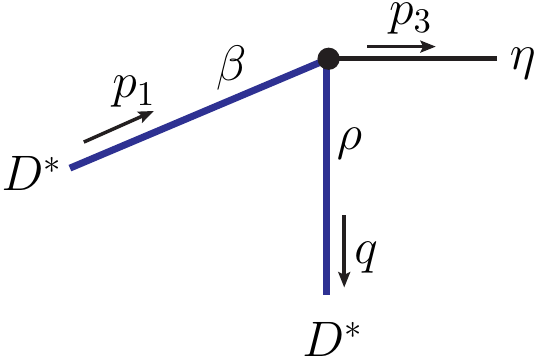}} &\widehat{=}& \frac{1}{2} g_{\mathcal{D}^\ast \mathcal{D}^\ast \mathcal{P}} \epsilon_{\rho \theta \tau \beta } (ip_3^\theta) (-ip_1^\tau-iq^\tau). \label{h1}
\end{eqnarray}
In addition, the involved propagators are given by
\begin{eqnarray}
\raisebox{-4pt}{\includegraphics[width=0.15%
\textwidth]{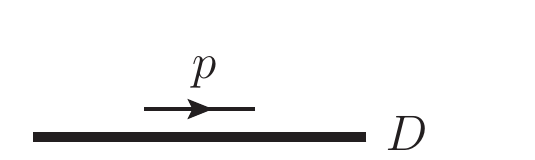}} &\widehat{=}& \frac{i}{p^2-m_D^2}, \label{Eq:propD}\\
%\end{eqnarray}
%\begin{eqnarray}
\raisebox{-4pt}{\includegraphics[width=0.15%
\textwidth]{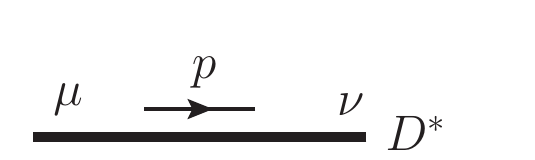}} &\widehat{=}& \frac{i(-g^{\mu \nu} +p^\mu p^\nu/m_{D^\ast}^2)}{p^2-m_{D^\ast}^2}.\label{Eq:propDStar}
\end{eqnarray}

\vfil
%\fi

\end{document}